\documentstyle [fleqn,11pt] {article}
\newcommand{\cd}{\cdot}

\newcommand{\al}{\alpha}

\newcommand{\de}{\delta}

\newcommand{\ep}{\epsilon}

\newcommand{\Ga}{\Gamma}

\newcommand{\la}{\lambda}

\newcommand{\si}{\sigma}

\newcommand{\ra}{\rightarrow}

\newcommand{\lap}{\triangle}

\newcommand{\be}{\begin{equation}}
\newcommand{\ee}{\end{equation}}
\newcommand{\bea}{\begin{eqnarray}}
\newcommand{\eea}{\end{eqnarray}}
\newcommand{\bean}{\begin{eqnarray*}}
\newcommand{\eean}{\end{eqnarray*}}
\newcommand{\dd}{\partial}

\begin{document}

\begin{titlepage}

\begin{flushright} ZU--TH9/94   \end{flushright}
 \vspace*{2cm}

\begin{center}{ \huge A new Contribution to Cosmological Perturbations
	of some Inflationary Models \vspace{1.5cm}}

  R. Durrer$^\#$ and  M. Sakellariadou$^*$\vspace{1cm}

$^\#$ Universit\"at Z\"urich, Institut f\"ur Theoretische Physik,\\
Winterthurerstrasse 190, CH-8057 Z\"urich, Switzerland
\vspace{0.5cm}\\
$^*$ Laboratoire de Gravitation et Cosmologie Relativistes, \\
 Universit\'e Pierre et Marie Curie, CNRS/URA 769,\\
Tour 22/12, Bo\^\i te 142,
4 place Jussieu, 75252 Paris Cedex 05, France.\\

\end{center}
\vspace{1.5cm}

\begin{abstract}
We show that there are inflationary models for which perturbations in the
energy momentum tensor, which are of second order in the scalar field,
 cannot be neglected.  We first specify the conditions under which the
usual first order perturbations are absent. We then analyze  classically,
the growth and decay of our new type of perturbations for one mode
of fluctuations $\de\phi_k$ in the scalar field. We generalize this
analysis, considering the contribution from the whole spectrum
of $\de\phi$ to a given wavelength of geometrical perturbations.
Finally, we discuss the evolution of the perturbations during the
subsequent radiation dominated era and discuss the resulting spectrum
of density fluctuations. In the case of a massless scalar field we find
a spectral index $n=4$. For massive scalar fields we obtain $n=0$ but
the resulting amplitude of fluctuations for inflation around a GUT scale
are by far too high. Hence, 'conventional' inflationary models must
not  be influenced by this new type of perturbations, in order to lead to
acceptable perturbations.
\vspace{1cm}
\end{abstract}
\end{titlepage}

\section{Introduction}

Inflation was originally proposed \cite{gu,sta} to solve some
shortcomings of  the standard cosmological model.
In addition, inflation can provide initial perturbations needed for
the formation of the structures observed in our universe.
Most inflationary scenarios
predict a scale--invariant spectrum of fluctuations \cite{bst}, known as
Harrison--Zel'dovich spectrum \cite{ha,ze}, which
 was found to be consistent
with  observations performed by  the COBE satellite \cite{sm,wr}.
This remarkable agreement motivated many of us in studying the evolution
of perturbations in the context of inflation. In particular, since
the amplitude of fluctuations poses the stringest constraints on
inflationary models , we are interested in analyzing the growth and decay
of cosmological perturbations.

To our knowledge, up to now the only perturbations of the energy momentum
tensor which have been considered, were first order in $\de\phi$.
(Recently we where told that quantum fluctuations which are second order
in $\de\phi$ have been investigated \cite{MS}.)
On the other hand, since successful inflation requires that $\dot{\phi}$
and $V,_\phi$ are
small enough, one can envisage situations where $\dot{\phi}$ is of the
order of $\de\dot{\phi}$ and $V,_{\phi}\ll V,_{\phi\phi}\de\phi$.
In this case, all
contributions to the perturbation of $T_ {\mu\nu}$ are second order in
$\de\phi$.
However, it has to be stressed that these perturbations
correspond to first order perturbations in both, $T_{\mu\nu}$ and  the
geometry, and thus they can be analyzed  by usual cosmological
perturbation theory.
This is the issue addressed in this paper.

Our work is organized as follows: In the next section, we analyze the
conditions on the potential for which first order perturbations can be
neglected. In section 3 we study the growth and decay of geometrical
fluctuations induced by a single mode perturbation in $\de\phi$. We
proceed in section 4 by generalizing to a generic spectrum of $\de\phi$.
In this case the whole spectrum contributes to the geometrical
perturbations
 of a given wavelength. In section~5, the evolution of perturbations is
followed through the radiation dominated era until second horizon crossing
to calculate the resulting power spectrum of density fluctuations.
We finally summarize our conclusions.

Since we are only concerned with the very early universe, we neglect
spatial curvature.
Furthermore, we use the signature $(-,+,+,+)$, such
that in conformal time $\eta$,
the metric is given by $ds^2=a^2 (-d\eta^2 +d{\bf x}^2)$.
We denote by prime and dot the
derivatives with respect to conformal and cosmological time respectively.
In what follows, $G$ is the Newtonian gravitational constant,
$m_{Pl}=1/\sqrt{G}$ is the Planck mass ($c=\hbar=1$) and
$H=\dot{a}/a=a'/a^2$ is the Hubble parameter.

\section{When are second order perturbations important?}

Let us consider a minimally coupled real scalar field $\phi$ with
Lagrangian density
\be
	L={1\over 2}\dd_\mu\phi\dd^\mu\phi -V(\phi).  \label{2L}
\ee
The stress energy tensor of $\phi$ is  given by
\be
	T_{\mu\nu}=\dd_\mu\phi \dd_\nu\phi -L g_{\mu\nu}. \label{2T}
\ee
During inflation, the energy density of the universe is dominated  by the
potential term $V$ and thus assuming a homogeneous and isotropic
inflaton field $\phi$, we have
$(1/2)\dot {\phi}^2 \ll V(\phi)$. The equation of motion
for $\phi(t)$
\be
\ddot{\phi} +3H\dot{\phi} +V,_{\phi}(\phi) = 0 \label{2EM}
\ee
(where $V,_{\phi}$ denotes the derivative
of $V$ with respect to $\phi$) and  Friedmann's
equation, $H^2= (8\pi G/3) V$, imply $|V,_{\phi}|
\ll \sqrt{48\pi}V/m_{Pl} $.

First order perturbations in the energy momentum tensor (\ref{2T}) have
the form
\be
\dot{\phi}\de\dot{\phi} ~~~ \mbox{ or }~~~  V,_{\phi}\de\phi . \label{21}
\ee
We now  want to study under which conditions the first order perturbations
in $\de\phi$ are negligible, namely, when
$\dot{\phi}\sim \de\dot{\phi}$ and
\be
   |V,_{\phi}|\ll \max\{|V,_{\phi\phi}\de\phi|, |{H\over a}\de\phi'|,
    {1\over a^2}|\de\phi''|, {1\over a^2}|\lap(\de\phi)|\} ~, \label{1}
\ee
where $\lap$ denotes the usual three dimensional Laplacian.
Let us consider the  scalar field to be of the form
\be
\phi=\phi_0+\de\phi , ~~~ \mbox{ with $\phi_0=$  constant and  }~~~
\de\phi= \de\phi_k=mA_k(\eta)e^{ikx}, \label{2f}
\ee
where we cast the dimension of $\de\phi$ in the mass parameter $m$, so that
$A_k$ is a dimensionless quantity. In the next section we will
appropriately choose the value of $m$. Provided  condition (\ref{1})
is met, we can then write the equation of motion for $\de\phi $ as
\be
A''_k + 2HaA'_k+A_k(k^2+a^2B)=0, \label{2A'}
\ee
where
\[B\equiv V,_{\phi\phi}(\phi_0).\]
We want to emphasize that the geometrical terms usually present in the
above equation (see, {\em e.g.} \cite{mfb}) do not contribute in our
case, up to first order, since $\dot{\phi}$
and $V,_\phi$ are already  first order perturbations.
The solution of (\ref{2A'}) is given by
\be
A_k=A_{in}(k)y^{3/2}J_{\pm \nu}(y), ~~~ \mbox{ with }~~~
	y\equiv {k\over Ha}
{}~~ \mbox{ and }~~~ \nu \equiv {3\over 2}\sqrt{1-{4B\over 9H^2}},
\label{2A}
\ee
where by $J_\al$ we denote the Bessel function of order $\al$.
As we will  show later,  for suitable initial conditions $A_{in}(k)$
is actually independent of $k$, so that it can be absorbed in the
definition of $m$. Thus, without loss of generality we set $A_{in}(k)=1$.
For the space and time derivatives of $\de\phi$ we then find
\bea
\de\dot{\phi}&=&-{k\over 2a}my^{1/2}[3J_{\pm \nu}(y)+yJ_{\pm \nu-1}(y) -
yJ_{\pm \nu +1}(y)]e^{ikx}, \label{2tdf} \\
\de\phi,_j &=& ik_jmy^{3/2}J_{\pm \nu}(y)e^{ikx}.\label{sdf}
\eea
Before proceeding with the discussion of the conditions imposed on the
form of
the potential $V$, we will distinguish between super-- and sub--horizon
perturbations. (Here, as usual, super (sub)-- horizon denote larger
(smaller) than the Hubble radius respectively.)

For super--horizon perturbations, $\la = a/k\gg 1/H$,
{\em i.e., } $y\ll 1$. For the least decaying mode we then find
\be
   |\de\dot{\phi}| = H|({3\over 2}-\nu)\de\phi| \label{11} .
\ee
For $B\ge 0$,
\[|\de\dot{\phi}^2|\le B |\de\phi|^2 \ll V(\phi_0),\]
where the last inequality sign has to hold in order for the universe
to be in
an inflationary phase. Therefore, if $|\de\dot{\phi}|$ was small initially,
it remains small during inflation for arbitrary positive values of $B$
(which is of course also intuitively clear). Furthermore, on super--horizon
scales $|\lap\de\phi |= k^2|\de\phi | \ll |\de\phi ''|$ and the last term
in (5) can thus be neglected.
The limit on $V,_\phi$ from (\ref{1}) then reduces to
$|V,_\phi(\phi_0)|\ll B|\de\phi|$. After $N$ e--foldings of inflation
$|\de\phi| \approx m|\exp[N(\nu-3/2)]|$, so finally the conditions
imposed on $V$ are
\bea
|V,_\phi(\phi_0)|\ll &Bm\exp(-3N/2), & \mbox { for ~} B> H^2 \label{c1}\\
|V,_\phi(\phi_0)|\ll &Bm\exp[-N({3\over 2}-\nu)], & \mbox { for ~}
	B< H^2  \label{c2}
\eea
On the other hand, for $B<0$, $A_k$ grows according to
$A_k \propto a^{\nu-3/2}$. For $\de\dot{\phi}$ not to grow extensively
during $N$ e--foldings of inflation, we hence have to require
\[ N(\nu-3/2) \le 1 ~~\mbox{ or, equivalently, }~~ -H^2/N<B  .\]
In that case the conditions on $V$ become
\be
	|V,_\phi(\phi_0)|\ll Bm ~~\mbox{ and }~~ B> {-H^2\over N}  .
\label{c3}
\ee

For sub--horizon perturbations, $\la =a/k\ll 1/H$,
{\em i.e., } $y\gg 1$, we have
\be
  |\de\dot{\phi}| \approx mHy^2 \approx Hy|\de\phi| = {k\over a} |\de\phi|
\ee
and
\be
 |\de\phi,_j|\approx k_j|\de\phi|\approx a|\de\dot\phi|.
\ee
Therefore, the conditions on the potential $V$ result in
\be
|V,_{\phi}(\phi_0)|\ll \max \{B|\de\phi|, {k^2\over a^2}|\de\phi|\}.
\label{c4}	\ee
On sub--horizon scales, classical perturbations decay and quantum effects
are important, and  a quantum analysis has to be performed.

Provided $0<B\ll H^2$, the $<(\de\phi)^2>$ can be approximated by
\be
<(\de\phi)^2>\approx {1\over (2\pi)^3}\int_0^{\infty}{1\over 2a^3
\sqrt{B+(k/a)^2}} d^3k + {H^2\over 4\pi^2}\int_o^{Ha}
({k\over Ha})^{2B/3H^2}{dk \over k}, \label{2ev}
\ee
which suggests that \cite{li}
\be
n_k\equiv {a^3H^2\sqrt{B+(k/a)^2}\over 2k^3}({k\over Ha})^{2B/3H^2}=
{\sqrt{1+B/(yH)^2}\over 2y^2}y^{2B/3H^2},
\ee
denotes the number density of $\phi$--particles.
 On sub--horizon scales, $n_k\ll 1$, which means that the quantum
analysis is needed. On the other hand, on super--horizon scales,
$n_k\gg 1$ and thus our classical analysis is indeed sufficient.

A quantum analysis study in the case of sub--horizon perturbations
shows that
the main contribution to (\ref{2ev}) is given by the second term, which
approximately yields
\be
\de\phi(k)\approx {H\over 2\pi}({k\over Ha})^{B/3H^2}\approx {H\over 2\pi}.
\ee
A natural choice for $m$, in the case that $B\ll H^2$, is
therefore $m\equiv H/(2\pi)$, leading to
\be
\de\phi({\bf x})=\int_0^{Ha}
{H\over 2\pi} e^{i{\bf k}\cdot {\bf x}} ({k\over Ha})^
{3/2}J_{\pm\nu}({k\over Ha})d^3k,
\ee
for super--horizon perturbations.

In the situation where the above imposed conditions on the potential $V$,
namely equations
(\ref{c1}), (\ref{c2}), (\ref{c3}) and (\ref{c4})
are satisfied, the perturbations
in $T_{\mu\nu}$ are given by
\be
\de T_{\mu\nu}=(\de\phi),_\mu (\de\phi),_\nu -{1\over 2}g_{\mu\nu}
(\de\phi),_\la (\de\phi),^\la +g_{\mu\nu}V,_{\phi\phi}\de\phi \de\phi.
\label{2dT}
\ee

In the following two sections, we will study the behavior of the
geometrical perturbations induced by the above $\de T_{\mu\nu}$.
Before proceeding with this study, we would like to briefly discuss
some inflationary models, for which our conditions are met.

\begin{itemize}
\item {\bf Old Inflation:} In old inflation \cite{gu},
the inflaton field $\phi$ is
sitting in a minimum of the potential $V$, so that $B>0$ and $V,_{\phi}=0$,
which means that our conditions are satisfied. However, since in this model
the phase transition is of first order and, thus, proceeds by bubble
nucleation, it has the problem of the so--called graceful
exit \cite{hms,gw}.
This means that the bubble nucleation process is slow with respect to the
universal expansion and, therefore, the phase transition can never be
completed. On the other hand,  if this problem is solved, ({\em e.g.}, by
extended or hyper--extended inflation \cite{st}) then the density
fluctuations
produced by bubble nucleation will, most probably, be the dominant ones.
\item {\bf New Inflation:} In new inflation \cite{lias},
the inflaton field $\phi$ is
sitting either in a minimum or in a maximum of the potential $V$,
depending on the details of the model. Nevertheless, if $B<0$, the slow
rollover condition requires $|B|$ to be small. Thus,  our
condition, namely, $B>-H^2/N$, may well be satisfied. In such a case,
one might think that the problems usually arising with first order
perturbations ({\em i.e.}, the requirement of extremely flat potential and
of a very weakly coupled scalar field) can be avoided. However, a more
detailed study has to be performed.
\item {\bf Supersymmetric Inflation:} In certain models of
 supersymmetric inflation \cite{hrr}, both $V,_{\phi}$ and $B$ vanish
and  our conditions are certainly satisfied. Thus our analysis can
indeed be applied.
\item {\bf Chaotic Inflation:} In chaotic inflation \cite{linde},
$V\propto \la \phi^n$, so that $V,_{\phi}\sim B\phi_0$ and therefore,
$V,_{\phi}\ll B\de\phi$ requires that $\phi_0\ll\de\phi$, which is of
course inconsistent with perturbation analysis. As a conclusion, within the
context of chaotic inflation our second order perturbations are
irrelevant.
\item {\bf de Sitter universe:} Regarding an arbitrary scalar field within
an inflationary model, our second order perturbation analysis can be
applied to study the fluctuations of that field, provided either the
potential is very flat, or, the field is sitting in a minimum of a
potential, which then can be of an arbitrary form.
\end{itemize}

\section{Perturbation analysis for one mode}

In this section we will investigate the evolution of second order
perturbations in the scalar field, which are of
 the specific form  $\de\phi =mA_k(\eta)e^{i{\bf k\cd x}}$.
Since the quantum analysis is only valid for $B\ll H^2$, we now restrict
ourselves to only this case.  As we have seen before,
the classical analysis applies only  for
super--horizon perturbations and  the value of $\de\phi$ at horizon
crossing is given by $\de\phi(y=1) = H/(2\pi)$. The solution,
Eq.\ (\ref{2A}),
of the classical equation of motion, Eq.\ (\ref{2A'}),  for
$\de\phi$ implies that the matching condition at horizon crossing,
 with $m\equiv H/(2\pi)$, yields
\[ A_{in}(k) = 1/J_{\pm\nu}(1) \sim 1~,\]
which is $k$--independent. This justifies to drop the prefactor
$A_{in}(k)$ in the sequel.
We now parametrize the  energy momentum tensor Eq.\ (\ref{2dT}) by
\bea
 \de T_{00} &=& m^2k^2f_\rho  ,\\
 \de T_{0j}  &=& m^2kf_v,_j  ,\\
  \de p &\equiv& {1\over 3}\de T_i^i = {m^2k^2\over a^2}f_p ,\\
\de\tau_{jl} &=& m^2k^2(f_\pi,_{jl}-{1\over 3}\lap f_\pi\de_{jl}) ,
\eea
where $\de\tau_{jl}$ is the traceless part of the perturbed stress tensor.
In terms of the function $A_k$ defined above, we obtain
\bea
f_\rho &=& {1\over 2k^2}[A_k^{'2} -A_k^2(k^2-a^2B)]e^{2i{\bf k\cdot x}},
  \label{fr} \\
f_p &=& {1\over 2k^2}[{1\over 3}k^2A_k^2+A'^2 -a^2BA_k^2]
	e^{2i{\bf k\cdot x}},
\label{fp} \\
f_v &=& {1\over 2k}A_kA'_ke^{2i{\bf k\cdot x}} , \label{fv}\\
f_\pi &=& A^2_ke^{2i{\bf k\cdot x}} . \label{fpi}
\eea

Using the perturbed Einstein equations, we will now relate
this energy momentum tensor to the perturbations in the geometry.
In the case of scalar perturbations, which are the relevant ones
for structure formation, geometrical perturbations can
be expressed in terms of the Bardeen potentials $\Phi,~ \Psi$  \cite{ba}.
In this paper  we only analyze scalar perturbations.

Let us now briefly discuss the physical meaning of the Bardeen potentials.
They were originally introduced because they  are invariant under
linearized coordinate  transformations, which, in the context
of cosmological perturbation theory, are called  gauge transformations.
As it has been shown \cite{ste,d93} (Stewart lemma),  gauge invariant
quantities are perturbations of tensor fields with either vanishing,
or constant, background
contribution.  Since homogeneous and isotropic universes
(Friedmann--Lema\^{\i}tre models) are conformally flat, they have a
vanishing
Weyl tensor. Thus, according to the Stewart  lemma, its perturbations are
gauge invariant. The Bardeen potentials are related to the electrical
part of the Weyl tensor by
\be E_{ij} \equiv a^{-2}C_{i0j0}={1\over 2}[ (\Phi-\Psi)_{|ij} -
	{1\over 3}\lap(\Phi-\Psi)] ~; \label{E}
\ee
the magnetic part vanishes for scalar perturbations
\cite{bde}.
The amplitude of perturbations in the geometry is typically given by
\[  {\cal A}\equiv \max_{\mu\nu\la\si\al\beta}
	\{|C_{\mu~\nu}^{~\la~\si}/R^\al_\beta\|\} \approx
	{k^2\over a^2H^2}|\Phi-\Psi| ~,\]
where $R^\al_\beta$ is the Ricci curvature. In the inflationary universe
\be {\cal A} \approx y^2(\Phi-\Psi)  ~. \label{AA}  \ee
Hence, the Bardeen potentials  determine the amplitude of perturbations at
horizon crossing. On super--horizon scales, where our analysis is
relevant, the amplitude of perturbations is constant only if the Bardeen
potentials grow as $a^2$.

Within the context of this paper, $\de T_\mu^\nu$
represents the perturbation of the constant tensor field
$V(\phi_0)\de_\mu^\nu$, thus it is gauge invariant and it can
be regarded  as a seed perturbation according to \cite{d90}.
The Bardeen potentials are then  given
by \cite{d90}
\bea
 \lap\Phi &=& -\ep[f_\rho k^2 + 3(a'/a)kf_v]  \label{phi}~; \\
 \lap(\Phi+\Psi) &=& -2\ep\lap f_\pi \label{phps} ~,
\eea
where $\ep \equiv 4\pi (m/m_{Pl})^2$  must be much smaller than 1 for
linear perturbation analysis to hold.
For $B\ll H^2$, $\ep $ becomes $\ep = H^2/(\pi m_{Pl}^2)$.
Equations (\ref{phi}) and (\ref{phps})
imply that in our case the Bardeen potentials are
\bea
 \Phi &=& {\ep\over 2k^2}
  \left[ A_k'^2 - A_k^2(k^2-a^2B) +3{a'\over a}A_kA'_k\right]
	e^{2i{\bf k\cdot x}}~;  \label{pot1}\\
 \Psi &=& -{\ep\over 2k^2}
  \left[ A_k'^2 + A_k^2(3k^2+a^2B) +3{a'\over a}A_kA'_k\right]
	e^{2i{\bf k\cdot x}} ~. \label{pot2}
\eea

On super--horizon scales ($y\ll 1$), $A_k$ given by Eq.\ (\ref{2A})
reduces to
\be
    A_k ={2^\nu \over \Gamma (1-\nu)}y^{3/2-\nu}~.\label{2Ak}
\ee
Inserting this result in equations (\ref{pot1}) and (\ref{pot2}), we find
\bea
   \Phi &=&{-2^{2\nu -1}~\ep\over [\Ga (1-\nu)]^2}y^{3-2\nu}
	e^{2i{\bf k\cdot x}}~; \\
   \Psi &=&{-3\cd 2^{2\nu -1}~\ep\over [\Ga (1-\nu)]^2}y^{3-2\nu}
  e^{2i{\bf k\cdot x}}~.
\eea
For  the relevant regime, $B\ll H^2$ and  the above expressions lead to
\be
  3\Phi=\Psi={-3 ~\ep \over \pi} e^{2i{\bf k\cdot x}}~.
 \label{Pinf} \ee
The Bardeen potentials are constant in time meaning that the amplitude
of curvature perturbations decays exponentially.

We shall see in Section~5, how the amplitude $\ep$ of $\Psi$ is connected
with observed fluctuations in the cosmic microwave background and what
its value has to be due to a normalization, e.g. according to the COBE
quadrupole.

\section{Perturbation analysis for the whole spectrum}

So far we have only considered the contribution of one wave
vector ${\bf k}$.
However, since the Bardeen potentials are quadratic in $\de\phi$, to a
given wave vector ${\bf k_1}$ of the ($\Phi, \Psi$)  spectrum, there is a
contribution from all wave vectors ${\bf k}$ of $\de\phi$. We therefore
address in this section, the contribution of the whole spectrum of
${\bf k}$ in the evolution of the Bardeen potentials.

We use the general ansatz
\[\de\phi=mA({\bf x},\eta).\]
The $f_{\rho}, f_v, f_{\pi}$ functions, which enter
in the Bardeen potentials, become
\bea
  f_{\pi}({\bf x}) &=&A^2~;\\
  \nabla _j  f_v({\bf x}) &=&lA'\nabla_j A~;\\
  f_{\rho}({\bf x}) &=&l^2[A'^2 +(\nabla A)^2 +a^2 B A^2]/2~,
\eea
where $l$ denotes an arbitrary length scale, for example $l=1/k_1$.
 To find the spectrum of the above functions we now perform a Fourier
transform. To obtain a finite result, we have to introduce a cutoff.
Taking into account that the fluctuations $A(k)$ are distributed with
uncorrelated phases, it makes sense to neglect contributions from
wavelengths
$\la<\la_1$, i.e., to choose the cutoff $k_1$ and correspondingly the
normalization volume $k_1^3$. Using the convolution theorem, we then obtain
\bea
f_{\pi}({\bf k_1}) &\approx & {1\over (2\pi)^{3/2} k_1^3}\int_0^{k_1} d^3k
                A(|{\bf k}|) A(|{\bf k_1} -{\bf k}|)~;\\
f_v ({\bf k_1}) &\approx &  {1\over (2\pi)^{3/2}k_1^4}\int_0^{k_1}
	d^3k      A'(|{\bf k}|) A(|{\bf k_1} -{\bf k}|) {\bf k_1} \cdot
({\bf k_1}-{\bf k})~;\\
f_{\rho}({\bf k_1}) &\approx &
       {1\over 2(2\pi)^{3/2} k_1^5}\int_0^{k_1} d^3k
              [A'(|{\bf k}|) A'(|{\bf k_1} -{\bf k}|)+\nonumber \\
  & &	({\bf k} \cdot
({\bf k_1}-{\bf k})+a^2B) A(|{\bf k}|) A(|{\bf k_1} -{\bf k}|)]~.
\eea

With
\[A(|{\bf k}|)={2^\nu \over \Gamma (1-\nu)}y^{3/2-\nu},\]
the above integrals finally yield
\bea
   f_{\pi}({\bf k_1}) &\approx & {\sqrt{8\over 9\pi^3}}~;\\
   f_v({\bf k_1}) &\approx & -{B\over 3H^2} \approx 0 ~;\\
   f_{\rho}({\bf k_1}) &\approx & {f_\pi}[{B^2\over 9H^4}
	+{3\over 5} +{B\over H^2}]  \approx{3f_\pi\over 5}
  ~, \eea
where $y_1 = k_1/(Ha)$.
Inserting these results in equations (\ref{phi} and
 \ref{phps}), we find for the Bardeen
potentials
\be
\Phi\approx -\Psi \approx \ep{\sqrt{2\pi}\over 5} ~.
\ee

This result shows that the Bardeen potentials remain constant during
the inflationary period and
the geometrical perturbation amplitude $\cal A$, introduced in the
last section, thus decays like $1/a^2$.
 However, the relevant physical quantity for structure
formation, is the change of $\de\rho/\rho$ between the first and
second horizon crossing. At horizon crossing, $\de\rho/\rho$ is well
approximated by the Bardeen potentials \cite{mfb}. Hence, we will  now
study the evolution of the Bardeen potentials during the radiation
dominated era, until the time of second horizon crossing.

\section{The radiation dominated era and second horizon crossing}
In this section $H$ without argument  denotes the value of the Hubble
parameter during inflation. The value of a quantity at the end of inflation
is indicated by the subscript $_{end}$ and at present time  by a
subscript $_0$.

\subsection{The Bardeen potentials at second horizon crossing}
According to the general equation of motion for $\de\phi$,
(\ref{2A'}), the evolution of the perturbation amplitude $A_k$ in the
radiation dominated regime is given by
\be {d^2A_k\over dy^2}+{2\over y}{dA_k\over dy} + (1+\al y^2)A_k
	= 0    \label{radA'} ~,
\ee
where $y=k/(aH(y))=k\eta$  in the radiation dominated epoch, and
$\al\equiv a^2B/(k^2\eta)^2 =const$. Since at the end of inflation
$\eta=1/(aH)$ to a very good accuracy, we find
\[ \al={B\over H^2y_{end}^4} ~.\]
Here $y_{end}$ is the value of $y$ at the end of inflation and $H$ denotes
the Hubble parameter during the inflationary period.
For $B=0$, (\ref{radA'}) is exactly solved in terms of  Bessel functions
and the 'growing mode' is given by
\[A_k(y) = {1\over \sqrt{y}}J_{1/2}(y) ~.\]
This solution is still approximately valid as long as $\al y^2<1$.
To require $\al y_{end}^2<1$ for all physically interesting scales,
$Ha_0\le k \le H_0a_0$ yields
in terms of the mass parameter $B$
\be B \ll z_{end}^2H_0^2  \sim (10^{-6}eV)^2\cd (T_{end}/10^{15}GeV)
\label{Bmin} ~. \ee
Obsevationally, such a  small mass can not be distinguished from
$B=0$. The general solution to (\ref{radA'}) at $y=1$ can be
found by exponentiating the matrix
\bean M(y,y_{end}) &=& \int_{y_{end}}^y\left(\begin{array}{cc} 0 & 1\\
-(1+\al y^2) & -2/y \end{array}\right) dy  \\
    &=& \left( \begin{array}{cc} 	0 & y-y_{end}\\
-(y-y_{end}+\al/3( y^3-y_{end}^3)) & 2\ln(y_{end}/y) ~ \end{array}\right),
\eean
which determines the  solution according to
\[ \left(\begin{array}{c} A(y)\\ A'(y)\end{array}\right) =
	\exp(M(y,y_{end}))
    \left(\begin{array}{c} A(y_{end})\\ A'(y_{end})\end{array}\right)~.\]
Unfortunately $M$ cannot be diagonalized; but explicit exponentiation
(by expansion in power series)
yields a simple, summable result if $|M_{21}|\gg  |M_{12}|,|M_{22}|$,
 which is valid for $\al \gg 1$ and $y\gg y_{end}$,
In the limit $\al\gg 1$ we find
\[ \exp(M(1,y_{end})) \approx (1+\ln(y_{end}))\left( \begin{array}{cc}
  \cos\sqrt{\al/3} & \sqrt{3/\al}\sin\sqrt{\al/3} \\
 -\sqrt{\al/3}\sin\sqrt{\al/3} & \cos\sqrt{\al/3} \end{array}
  \right) ~. \]
For each element in $\exp(M)$ we have kept only the highest power in $\al$.
Using the exact solution during the inflationary epoch, eqn. (\ref{2Ak}),
which yields
\be
    A'(y_{end})= {\al\over 3}y_{end}^3A(y_{end}) ~, \label{A'end}
\ee
 we obtain for $A_k, A'_k$ at $y=1$
\bean A_k(1) &=& (1+\ln(y_{end}))[ \cos\sqrt{\al/3}+
	 y_{end}^3\sqrt{\al/3}\sin\sqrt{\al/3} ] A_k(y_{end})\\
   A'_k(1) &=& (1+\ln(y_{end}))[(\al/3)
	 y_{end}^3\cos\sqrt{\al/3}
	- \sqrt{\al/3}\sin\sqrt{\al/3} ] A_k(y_{end})  ~.
\eean
In the radiation dominated era eqn. (\ref{pot2}) for the Bardeen potential
$\Psi$ yields
\be \Psi_{rad}(y) = -(\ep/2)[ ({dA\over dy})^2 +A^2(3+\al y^2) +
	(3/y)A{dA\over dy}]  ~ . \label{Prad} \ee
For $\al\gg 1$ (which implies $y_{end}\ll 1$) this yields
\be \Psi_{rad}(y=1) \approx -{\ep\al\over 3} A^2(y_{end})
  ~ . \label{Prad1} \ee
Equation (\ref{Prad}) at $y=y_{end}$ with the initial condition
(\ref{A'end}) leads for $\al y_{end}^2\gg 1$ to
\[ \Psi_{rad}(y_{end}) \approx  -\ep\al y_{end}^2A^2(y_{end})
	 \propto y_{end}^{-2}	~. \]
On the other hand, from the inflationary universe calculation we have
$\Psi_{inf}(y_{end})=\ep =$ const. Assuming now, that the Bardeen
potentials do not jump when the universe enters the radiation dominated
phase, $\Psi$ has changed since the end of inflation until second
horizon crossing by a factor
\be
   	g(k) = \Psi_{rad}(1)/\Psi_{rad}(y_{end}) \approx y_{end}^{-2}/3 ~ .
\ee
This yields for the final Bardeen potential at second horizon crossing
\be \Psi(\eta_k) \approx -{\ep\over 3 y_{end}^2} =
	-{\ep(Ha_{end})^2\over 3k^2} ~~
	\mbox{ for }~~ \al =B/(H^2y_{end}^4)=BH^2a_{end}^4/k^4 \gg 1
	~ , \label{Psi1}\ee
where $\eta_k$ denotes the time of second horizon crossing of wave
number $k$, {\it i.e.} $\eta_k =1/k$ and $\ep=H^2/(\pi m_{Pl}^2)$.
In the case $B=0$ we find $\Psi_{rad}\approx$ const., independent of
$y_{end}$. This leads to a growth factor of $g(k)\approx 1$, and thus
\be \Psi(\eta_k) \approx	 -\ep ~~~ \mbox{ for }~~ B=0 ~ .
\label{Psi0} \ee
In the intermediate regime, ${\cal O}(\al)\sim 1$, we have solved
(\ref{radA'}) numerically. The solutions $A_k$ for some values
of $\al$ are represented in Fig.~1.

In Fig.~2 we  plot the value of $\Psi(\eta_k)$
 in units of $\ep$ as a function of physical wave number $k/a$ in units
of $H$ (In other words, we  plot $\Psi$ as a function
of $y_{end}$.). We  chose the parameter value
$B/H^2=0.01$. For small wave numbers, $\al\gg 1$, the amplitude of $\Psi$
decays like $k^{-2}$ until approximately
$\al y_{end}^2=BH^2/(k/a_{end})^2 \sim 1$. At $(k/a_{end})^4 \sim BH^2 $,
{\it i.e.}  $ \al \sim 1$ the potential bends over to a constant.

This results remain qualitatively valid also in the matter dominated era
which we thus shall not treat seperately.

\subsection{The power spectrum}
Let us now briefly clarify what power spectrum results from the findings
of the last subsection. Since the  explanations of  spectral index
and  scale invariance found  in the literature are often
somewhat confusing, we shall be rather explicit in this paragraph.

The power spectrum is defined by
\[ P(k,\eta) = |\de_k(\eta)|^2 ~, \]
 where  $\de_k(\eta)$ is the Fourier transform of the density fluctuation
$(\de\rho/\rho)(x,\eta)$. Clearly for perturbations which are larger than
the size of the horizon, $k<1/\eta$, $P(k,\eta)$ is gauge dependent. But
we are mainly interested in $P(k,\eta_0)$ on scales $k>1/\eta_0$ and
$a_0/k \approx \la_0 \ge 10$Mpc, {\it i.e.}, scales which are small
enough to be observable
today and large enough, so that they are probably not severely affected by
nonlinear clustering. The spectrum $P(k,\eta_0)$ is called scale invariant,
or a Harrizon--Zel'dovich spectrum, if the variance of the mass
perturbation on horizon scale $l_H=a\eta$ is constant, time independent:
\[ \langle({\de M\over M})^2_{l_H}\rangle = \mbox{ const.} \]
Here $\langle~ \rangle$ denotes statistical average over many
realizations of  perturbed
universes with identical statistical properties, and we assume that it can
be replaced by a spatial average.
\[ (\de M)_{l_H} = \int_{V_H}\de\rho d^3x =
	\rho\int_{V_H}(\de\rho/\rho)d^3x
	\approx \rho V_H\int_0^{k_H}\de_ke^{ikx}d^3k ~ . \]
Here we have assumed that perturbations on scales smaller than $l_H$
($k>k_H=1/\eta$) average to zero due to the integration over $V_H$,
and that perturbations on scales larger than $l_H$ ($k<k_H$) are
approximately constant in $V_H$, such that integration over $V_H$ just
gives rise to a factor $V_H$. Using $M=\rho V_H$, we obtain
\be
 (\de M/M )_{l_H} 	\approx \int_0^{k_H}\de_ke^{ikx}d^3k ~ ,
	\label{MH} \ee
such that
\bean
  \langle (\de M/M )^2_{l_H} \rangle	&\approx & \int d^3x
   \int_0^{k_H}d^3k\int_0^{k_H}d^3k'\de_k\de_{k'}e^{i(k+k')x}  \\
 &=& \int_0^{k_H}|\de_k|^2d^3k =\int_0^{k_H}P(k,\eta)d^3k ~ ,
\eean
For the first equal sign we have evaluated the spatial integration
leading to a $\de$--distribution and the $k'$ integration using
$\de_{-k}=\de_k^*$. If we now assume $P(k,\eta)$ behaves like a power
law in $k$, $P(k,\eta)\propto k^m$, scale invariance requires $m=-3$, such
that $\de_k(\eta_k) \propto k^{-3/2}$, where as before $\eta_k$ denotes
the conformal time at which the scale $k^{-1}$ enters the horizon,
{\it i.e.} $\eta_k= k^{-1}$. (We should not be concerned about the
logarithmic singularity for $k \ra 0$. We anyway have to introduce a
low end cutoff to (\ref{MH}) at  about $k_{H_0}$, since perturbations
on scales larger than the present Hubble radius cannot be distinguished
from background contributions.)

For a scale invariant spectrum we thus have to require
\[ \de_k(\eta) \propto k^{-3/2} ~ \mbox{ on super horizon scales.} \]
Since we always can choose a gauge such that $\de_k(\eta) \approx
 (\eta k)^2 \Psi_k(\eta)$, this yields
\[  \Psi_k(\eta_k)  \approx \de_k(\eta_k) \propto k^{-3/2} ~.\]
As is well known from linear perturbation theory (see e.g. \cite{Pee}),
matter perturbations on scales which enter the horizon during the
matter dominated era subsequently start growing according to
$\de_k(\eta) \propto a(\eta) \propto \eta^2$. Perturbations on smaller
scales which enter the horizon already in the radiation dominated era,
remain approximately constant until the time of equal matter and radiation,
$\eta_{eq}$, and then also grow according to $\de\propto a$.
We therefore end up with the following behavior of density fluctuations
at late times $\eta>\eta_{eq}$:

\[ \de_k(\eta)\approx \left\{  \begin{array}{cc}
      Dk^{1/2}\eta^2  & \mbox{ for } k^{-1} >\eta_{eq} \\
      Dk^{-3/2}(\eta/\eta_{eq})^2    & \mbox{ for } k^{-1} <\eta_{eq},
\end{array} \right.
\]
or
\[ P(\eta,k) \approx {D^2k\eta^4\over (1+(k/k_{eq})^2)^2} ~,~~
   \mbox{ where  }~ k_{eq}^{-1} =\eta_{eq}. \]
$D$ is a constant of proportionality which is determined by the value
of $\de_k$ at horizon crossing, $\de_k(\eta_k)=Dk^{-3/2}$.
On large scale we then find
\[ P(\eta_0,k) \propto k^n ~,~~~ \mbox{  with } n=1. \]
The power $n$ in this law is the so called spectral index.

 We have thus seen, a power spectrum $P(\eta_0,k) \propto k^n$ on
large scales leads to the behavior
\be \Psi(\eta_k,k) \propto k^{n/2-2} ~. \ee
Therefore, the behavior of $\Psi(\eta_k,k)$ in our model yields
spectral indices
\be n =\left\{ \begin{array}{ll}
      4  & \mbox{ for } B=0 \\
     0   & \mbox{ for } B\gg (10Mpc)^{-2} \approx (10^{-30}eV)^2,
\end{array} \right.   \label{spec}
\ee

As is discussed in detail in \cite{Sc}, the COBE experiment is still
compatible with  spectral indices within the bounds $-2\le n\le 2$,
with a peak in the likelyhood around  $n\approx 1.3$.
The case $B=0$ is therefore excluded by the COBE DMR limits on the
spectral index. The possibility $B\neq 0$ cannot
be ruled out from the COBE data alone. The relatively
strong clustering of galaxies on large scales (see e.g. \cite{Vo})
even strongly disfavors a large spectral index, but this always depends on
the matter model assumed (CDM is only consistent with $n\le 1$, whereas
HDM prefers larger spectral indices).

Let us now investigate the amplitude of the induced perturbations.
The value of $\ep=H^2/(\pi m_{Pl}^2)$ can be related to
the amplitude of the quadrupole of  microwave background anisotropies.
$C_2$ can quite generally determined by the density perturbations today
according to (see \cite{efs})
\be C_2\approx \int_0^\infty |\de_k(\eta_0)|^2(j_2(y)/y)^2dy ~, \ee
with $y=k\eta_0$.
Using $\de_k(\eta_0)=\ep y_{end}^{-2}(\eta_0 k)^2/3 =
(\ep/3) (\eta_0/\eta_{end})^2$
for large enough scales which are relavant for the quadrupole anisotropy,
we obtain in the case $B\neq 0$
\be C_2 \approx \ep^2{z_{end}^4\over 3000z_{eq}^2}
  ~. \ee
With the COBE quadrupole of $C_2 \approx 3\cd 10^{-11}$ this yields
$H\approx 0.02GeV$. A scalar field mass of $m =\sqrt{B}\ll H$,
 certainly incompatible with the standard model of particle
physics which is well tested at these low energies.

For $B=0$ one finds $C_2 \approx \ep^2 = (H/m_{pl})^2$, leading to
$H\approx 10^{16}GeV$, a typical GUT scale, but this possibility is
already ruled out due to its large spectral index.
Certainly, as soon as reliable intermediate measurements of the cosmic
microwave background anisotropies are available, the spectral index
$n$ will become even more constrained.

\section{Conclusions}

In this work we have investigated inflationary models in which the
perturbations of second order in $\de\phi$ are the dominant ones.
We have studied the evolution of these perturbations.
Depending on the mass of the scalar field, we found that the resulting
density fluctuations have  spectral index $n=4$ for massless
fields and  $n=0$ for massive scalar fields, the second of which is
compatible with the COBE DMR results.
However, the quadrupole amplitudes for the two cases  are completely
different:
\be C_2 \approx \left\{\begin{array}{ll}
	(H/1GeV)^6 & \mbox{ for } B \neq 0 \\
	(H/m_{pl})^2 & \mbox{ for } B= 0~. \end{array} \right. \ee
In the case $B\neq 0$, for an infationary model to yield small enough
perturbations,  one therefore must make sure that
the perturbations of second order in the scalar field discussed in
this work are substantially suppressed. Otherwise, due to the enormous
increase $\propto y_{end}^{-2}$ of the Bardeen potentials
during the radiation era, they will soon dominate
and spoil the perturbation theory.

For 'intermediate' masses, $m=\sqrt{B} \sim 10^{-4}eV$, the spectral
index can turn out to be of order $n \sim 0 - 2$ and the amplitude
$C_2\sim \ep^2$. (When the relevant physical scales,
$100 Mpc \le k^{-1} \le 1000Mpc$ happen to lay at the bending point of
$\Psi$ shown in Fig.~2.) But this requires severe fine tuning of the
scalar field mass and is thus not investigated any further.

We have not followed through the reheating period, the transition from
the inflationary to the radiation dominated phase. We  just assumed that
perturbations induced during this short time can be neglected on super
horizon scales, such
that the Bardeen potentials  pass continuously over the transition.
 If our scalar field represents the inflaton field, this assumption has
to be justified by further work. But if we   regard it as a scalar
field in an external de Sitter universe which later becomes radiation
dominated, our analysis is certainly valid.

In the radiation  era we have only analyzed one mode of
perturbations of the scalar field.  In principle we would have to perform
the convolution presented in Section~4 to obtain the entire contribution
to a given mode $k_1$ of the Bardeen potentials which are quadratic in
the scalar field. Repeating the calculation outlined in Section~4, one
immediately sees if $A_k~\sim const.$ (which is the case
for $B=0$) also the Bardeen potentials
turn out to be constant and of the same order of magnitude as for one
mode. On the other hand, for $B\neq 0$  $A_k$ is a rapidly oscillating
function of $\sqrt{\al} \propto 1/k^2$ and thus the main contribution
comes from the  wave number pair $k=k_1/2$ and we do not expect the
behavior of the spectrum to differ substantially from the one mode result.
Therefore, we expect the results (\ref{spec}) and (63) obtained in a one
mode analysis to  remain  valid for the full spectrum of fluctuations.
\vspace{0.6cm}\\
{\Large Acknowledgments} \vspace{0.3cm}

It is a pleasure to thank Robert Brandenberger, Nathalie Deruelle
and Slava Mukhanov, for a number of stimulating discussions. One of us,
M. S., would like to thank the Institut f\"ur Theoretische Physik
of Z\"urich University, for  hospitality during the preparation
of this work.

\newpage

{\Large\bf Figure Captions}
\vspace{1cm}\\
{\bf Fig. 1}\hspace{0.3cm} The amplitude of the scalar field, $A_k$
is shown as a function of $y$ for 3 characteristic values of $\al$.\\
Solid line: $\al=10^5$, with $B/H^2=10^{-4}$, such that
$y_{end}=0.056$.\\
Dotted line: $\al=120$, with $B/H^2=10^{-4}$, such that
$y_{end}=0.03$.\\
Dashed line: $\al=1$, with $B/H^2=10^{-4}$, such that $y_{end}=0.1$.
\vspace{0.5cm}\\
{\bf Fig. 2}\hspace{0.3cm} The amplitude of the Bardeen potential $\Psi$
at second horizon crossing, $k=\eta^{-1}$ is shown in arbitrary units as
function of $y_{end}=k/(a_{end}H)$. The dashed line shows the $k^{-2}$
behavior. In this example $B/H^2=10^{-4}$, so that $\al y_{end}^2=1$ for
$y_{end}=0.01$. The calculated curve even becomes steeper at somewhat
larger values of  $\al y^2$ and then, around $\al=1$ (corresponding to
$y_{end}=0.1$), bends over to a constant.

\end{document}